\documentclass[numberedappendix]{emulateapj}
\usepackage{natbib}
\newcommand{\Msun}{{\rm M_{\odot}}}

\newcommand{\kmps}{\, {\rm km \, s^{-1}}}

\newcommand{\pcc}{\,{\rm cm}^{-3}}

\slugcomment{}

\shorttitle{CO $J=2-1/1-0$ in M51}
\shortauthors{Koda et al.}

\begin{document}

\title{Physical Conditions in Molecular Clouds in the Arm and Interarm Regions of M51}

\author{Jin Koda\altaffilmark{1},
Nick Scoville\altaffilmark{2},
Tetsuo Hasegawa\altaffilmark{3},
Daniela Calzetti\altaffilmark{4},
Jennifer Donovan Meyer\altaffilmark{1},
Fumi Egusa\altaffilmark{5},
Robert Kennicutt\altaffilmark{6},
Nario Kuno\altaffilmark{7, 8},
Melissa Louie\altaffilmark{1},
Rieko Momose\altaffilmark{9},
Tsuyoshi Sawada\altaffilmark{3, 10},
Kazuo Sorai\altaffilmark{11},
Michiko Umei\altaffilmark{11}}
\altaffiltext{1}{Department of Physics and Astronomy, Stony Brook University, Stony Brook, NY 11794-3800}
\altaffiltext{2}{Department of Astronomy, California Institute of Technology, Pasadena, CA 91125}
\altaffiltext{3}{National Astronomical Observatory of Japan, NAOJ Chile Observatory, Joaqu\'{\i}n Montero 3000 Oficina 702, Vitacura, Santiago 763-0409, Chile}
\altaffiltext{4}{Department of Astronomy, University of Massachusetts, Amherst, MA 01003}
\altaffiltext{5}{Department of Space Astronomy and Astrophysics, Institute of Space and Astronautical Science, Japan Aerospace Exploration Agency, Japan}
\altaffiltext{6}{Institute of Astronomy, University of Cambridge, Cambridge CB3 0HA, United Kingdom}
\altaffiltext{7}{Nobeyama Radio Observatory, Minamimaki, Minamisaku, Nagano, 384-1305, Japan}
\altaffiltext{8}{The Graduate University for Advanced Studies (SOKENDAI), 2-21-1 Osawa, Mitaka, Tokyo 181-0015, Japan}
\altaffiltext{9}{Institute for Cosmic Ray Research, University of Tokyo, 5-1-5 Kashiwa-no-Ha, Kashiwa City, Chiba, 277-8582, Japan}
\altaffiltext{10}{Joint ALMA Office, Alonso de C\'{o}rdova 3107, Vitacura, Santiago 763-0355, Chile}
\altaffiltext{11}{Department of Physics/Department of Cosmosciences, Hokkaido University, Kita-ku, Sapporo 060-0810, Japan}

\email{jin.koda@stonybrook.edu}

\begin{abstract}
We report systematic variations in the emission line ratio of the CO $J=2-1$ and $J=1-0$ transitions ($R_{2-1/1-0}$)
in the grand-design spiral galaxy M51. The $R_{2-1/1-0}$ ratio shows clear evidence
for the evolution of molecular gas from the upstream interarm regions, passage into the spiral arms and
back into the downstream interarm regions. In the interarm regions,
$R_{2-1/1-0}$ is typically $<0.7$ (and often 0.4-0.6); this is similar to the ratios
observed in Galactic giant molecular clouds (GMCs)
with low far infrared luminosities. However, the ratio rises to $>0.7$ (often 0.8-1.0)
in the spiral arms, particularly at the leading (downstream) edge of the molecular arms.
These trends are similar to those seen in Galactic GMCs with OB star formation
(presumably in the Galactic spiral arms).
$R_{2-1/1-0}$ is also high, $\sim 0.8-1.0$, in the central region of M51. Analysis of the molecular excitation using 
a Large Velocity Gradient radiative transfer calculation provides insight into the changes in
the physical conditions of molecular gas between the arm and interarm regions:  cold and low density gas ($\lesssim 10$ K, $\lesssim 300 \pcc$) 
is required for the interarm GMCs but this gas must become warmer and/or denser in the more active star forming spiral arms.
The ratio $R_{2-1/1-0}$ is higher in areas of high $24\mu$m dust surface brightness
(which is an approximate tracer of star formation rate surface density) and high CO(1-0) integrated intensity
(i.e., a well-calibrated tracer of  total molecular gas surface density).
The systematic enhancement of the CO(2-1) line relative to CO(1-0) in
luminous star forming regions suggests that some caution is needed when using  CO(2-1) as a tracer of bulk molecular gas mass,
especially when galactic structures are resolved.

\end{abstract}


\section{Introduction}

The standard, albeit simplistic, picture of interstellar matter (ISM) phases posits that
giant molecular clouds (GMCs) are assembled in spiral arm shocks from
diffuse interarm HI gas and then photo-dissociated back into the atomic phase
by OB star formation within the spiral arms.
This picture predicts a rapid gas-phase change across spiral arms Ðfrom atomic
to molecular and back into the atomic after spiral arm passage.
New CO(1-0) observations of M51 suggest a very different picture in which there is little (or no)
gas-phase change between interarm and arm regions with the majority, 70-80\%, of the neutral gas
remaining molecular from arm entry through the inter-arm region and
into the next spiral arm passage \citep{Koda:2009wd}.
The molecular gas fraction could be even higher if dark H$_2$ gas exists
\citep[e.g., Planck Collaboration 2011; ][]{Grenier:2005lr}, which is not detectable in CO emission.
In this new picture, the obvious molecular spiral arms are assembled from pre-existing small GMCs
by spiral arm forcing and orbit crowding within the arms. A fraction of the gas forms dense cores in
the molecular spiral arms, leading to star formation \citep{Egusa:2011dq},
but the majority is shredded apart by spiral shearing motions and goes back to the interarm
regions as smaller GMCs \citep{Koda:2009wd, Wada:2004fu}.

Questions still remain as to what the physical conditions of the abundant
molecular gas in the interarm regions are and how these conditions change within the spiral arms.
The lowest rotational transition $J=1-0$ of the carbon monoxide (CO) molecule is
the most reliable tracer of the overall molecular gas mass and surface density in galaxies.
The transition energy of the CO(1-0) corresponds to 
$E/k\sim5.5$ K and this is conveniently below the typical observed temperatures of resolved GMCs \citep[$\sim10$ K; ][]{Scoville:1987vo};
in addition, the critical density for CO(1-0) collisional excitation in the presence of photon trapping
s a few $\times 100\, \rm\, H_2\, \pcc$
which is roughly the same as the average density within the GMCs \citep{Solomon:1987pr}.
Therefore, CO(1-0) line emission arises from the levels which are close to thermalized,
and the CO (1-0) luminosity 
traces their bulk molecular mass
\citep[see][]{Scoville:1987vo}.
Empirically, CO(1-0) luminosities of GMCs are correlated with their virial mass estimates.
A nearly constant mass-to-light ratio has been calibrated with virial masses of
resolved GMCs in the Galaxy \citep{Solomon:1987pr, Scoville:1987vo}
and in nearby spiral galaxies \citep{Bolatto:2008nz,Donovan-Meyer:2012ve}.
[Note that, despite the constant ratio observed in normal spiral galaxies,
there are discussions on its variations in low-metallicity dwarf galaxies
\citep{Israel:1997lr, Leroy:2009cy}, where self-shielding of GMCs from
ambient ultraviolet radiation may not be as effective.]

Within the GMCs only a small fraction of the gas is typically within the dense core regions leading to star formation. 
HCN(1-0), having a higher critical density \citep[$\sim 10^5\pcc$; ][]{Kohno:1996jv} than low-J CO lines,
traces the emission from such dense cores and this line emission exhibits a linear correlation with infrared (IR) emission, a tracer of embedded star formation,
over 7-8 orders of magnitude --
from starburst galaxies down to small dense cores within Galactic GMCs \citep{Gao:2004kg,Wu:2005tw}.
Recently, CO(3-2) has also been used to trace dense cores \citep{Muraoka:2007yq, Wilson:2009fk}.
Its higher critical density ($\gtrsim 10^4\pcc$) and relatively high temperature corresponding to
the upper energy level ($\sim33$ K) imply that it  traces the environments with active star formation
-- either dense star-forming molecular cores or gas heated by recent star formation.
This has been demonstrated by \citet{Komugi:2007vn} who found a linear correlation between surface densities of
CO(3-2) emission and star formation in nearby galaxies.

CO(2-1) emission has an upper-level energy temperature of $\sim16.5$ K and a collisional critical density
of $\sim 10^{3-4}\pcc$, slightly above the standard GMC temperature and density;
hence the CO(2-1)/CO(1-0) line ratio ($R_{2-1/1-0}$) can
 trace the change of physical conditions of
the bulk molecular gas.
In fact, the ratio shows a systematic variation in the Galaxy \citep{Sakamoto:1999rt}
and has been used to investigate the varying physical conditions \citep{Sakamoto:1997ys, Sawada:2001lr}.
$R_{2-1/1-0}$ has also been used to infer gas physical conditions in external galaxies,
especially in their central regions \citep{Knapp:1980uq, Braine:1992lr, Turner:1993fj, Aalto:1995lr, Harrison:1999kx}.

Attempts to map $R_{2-1/1-0}$ across galaxy disks have remained inconclusive.
\citet{Garcia-Burillo:1993ys} analyzed their 2-1 and 1-0 imaging of M51 and found no systematic variation in $R_{2-1/1-0}$
between the spiral arm and interarm regions.
However, the early CO(1-0) data \citep{Nakai:1994ec} used in their analysis appear to be
inconsistent by a factor of $\sim 2$ in flux with four other later measurements \citep{Koda:2011nx},
perhaps due to calibration difficulties 20 years ago.
\citet{Schinnerer:2010kr} analyzed $R_{2-1/1-0}$, as well as other molecular lines,
in the western spiral arm of M51 using data obtained with interferometers, which was
unfortunately combined with the old \citet{Nakai:1994ec} single-dish data with the apparent calibration problem.
Therefore, the question of large-scale variation of $R_{2-1/1-0}$ remains unanswered.
Here, we present new CO(2-1) and CO(1-0) imaging with a new and consistent observation technique 
and careful calibration for both lines.
These new data permit much-improved measurements of the spatial variations in $R_{2-1/1-0}$ in both arm and  interarm regions of M51.

\begin{figure*}
\epsscale{1.1}
\plotone{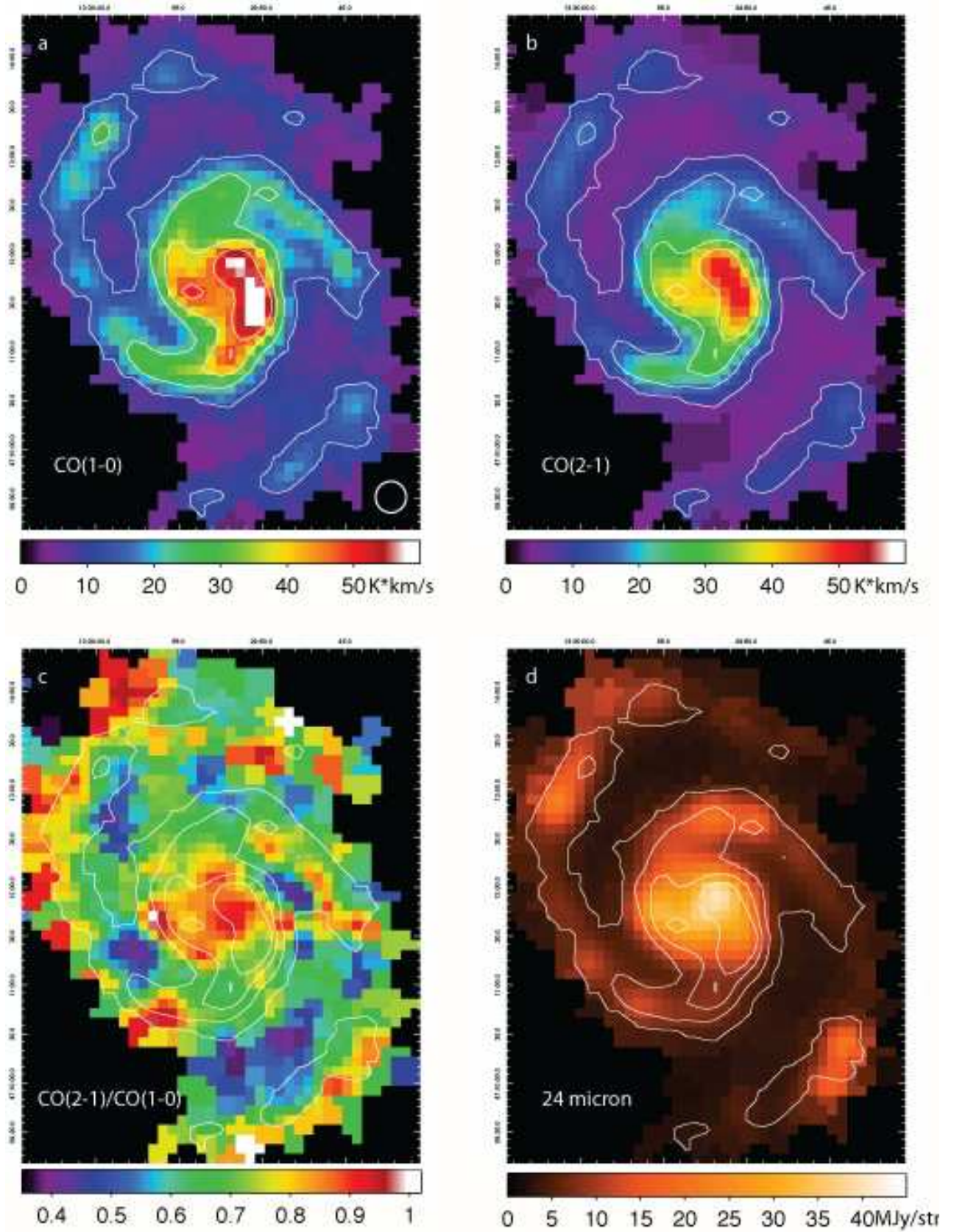}
\caption{Maps of the central $5.4\arcmin \times4.1\arcmin$ region of M51 in the J2000 coordinate system.
(a) CO $J=1-0$ integrated intensity, $I_{\rm CO(1-0)}$, map in units of main beam temperature.
(b) CO $J=2-1$ integrated intensity, $I_{\rm CO(2-1)}$, map in units of main beam temperature.
(c) CO $J=2-1/1-0$ line ratio, $R_{2-1/1-0}$ ($\equiv I_{\rm CO(2-1)}/I_{\rm CO(1-0)}$), map.
(d) Spitzer $24\mu m$ surface brightness, $f_{24\mu m}$, image.
The signal to noise ratios are high across the entire area including at the edges, i.e.,
$\gtrsim 10\sigma$ in both CO(1-0) and CO(2-1).
The circle at the bottom-right corner in (a) is the beam size of the CO $J=1-0$ map,
$19\arcsec.7$ ($\sim 780$ pc at the distance of 8.2 Mpc).
All other images are smoothed to this resolution.
Contours in all panels are at $I_{\rm CO(1-0)}=$ 12, 24, 36, and $48\,\rm K\cdot \kmps$.
The line ratio varies systematically between the spiral arms and interarm regions.
\label{fig:maps}}
\end{figure*}

\section{CO and Infrared Data}

The CO $J=1-0$ data are taken from \citet{Koda:2011nx}, which were taken as a part
of the CARMA NObeyama Nearby-galaxies (CANON) survey.
The original observations include both interferometer data from the Combined Array for Research
in Millimeter Astronomy (CARMA) and single-dish data from the Nobeyama Radio Observatory
45m telescope (NRO45).
However, we use only the NRO45 data here to avoid unnecessary concerns, if any,
about potential imaging artifacts introduced in the process of image reconstruction.
CARMA data, observing small angular scales, do not play a role in our analysis,
since we use a spatial resolution of $19.7\arcsec$.

The improved accuracy of the new data is due to the 25-BEam Array Receiver System
\citep[BEARS; ][]{Sunada:2000fk} and the On-The-Fly (OTF) mapping technique \citep{Sawada:2008xz}.
They enable us to map the large galaxy within a short observing duration, ensuring consistency
in calibration.
Multiple OTF scans in the RA and DEC directions are averaged to minimize systematic errors
which appear dominantly along the scan directions (note that this type of error could not be characterized
without the OTF technique). Therefore, the new data provide a much higher accuracy
in relative calibration across the map.
The absolute flux calibration is consistent with other independent measurements at a few percent level.
The details of the data reduction are given in \citet{Koda:2011nx}.
The final data cube has a spatial resolution of $19.7\arcsec$ (degraded by gridding from
the beam size of $15\arcsec$) and a $1\sigma$ noise of
$\sim 37$ mK on the main beam temperature scale in a 10$\kmps$ channel
at a pixel size of $5.96\arcsec \times 5.96\arcsec$.
It covers the entire molecular disks of NGC 5194 (main galaxy) and NGC 5195 (companion).

The CO $J=2-1$ data are from \citet{Schuster:2007zr}, which are distributed as a part of
the HERA CO Line Extragalactic Survey \citep[HERACLES; ][]{Leroy:2009zv}.
The data are taken with the 18 element focal plane heterodyne receiver array (HERA)
on the IRAM-30m telescope.
The OTF technique was also employed.
The 2-1 data cover the entire disk of NGC 5194, but do not include NGC 5195.
The original integrated intensity image is in \citet{Schuster:2007zr}.
The spatial resolution is $11\arcsec$, and the noise is $\sim 22$ mK in a 10$\kmps$ channel
on the main beam temperature scale. The CO(2-1) data cube was smoothed to the $19.7\arcsec$ resolution and regridded 
to the CO(1-0) reference grid. 
When comparing the two maps, we noticed that the global CO(2-1) distribution is offset by $\sim 1$ pixel
in the south-west direction with respect to CO(1-0). We therefore shifted the CO(2-1) data correspondingly
before making the integrated intensity maps with both lines integrated over the same velocity ranges referenced to
the observed rotational velocity field in the galaxy. This treatment does not change the conclusions in this paper,
as the spiral arms and interarm regions are considerably wider than one pixel.

The Spitzer 24$\mu m$ image and H$\alpha$ image are taken from the archive of
the Spitzer Infrared Nearby Galaxies Survey \citep[SINGS; ][]{Kennicutt:2003fd}.
The GALEX far-ultraviolet (FUV; $1539\AA$) image \citep{Gil-de-Paz:2007lj} is taken through GalexView.
We adopt the central coordinates of $(\alpha, \delta)_{\rm J2000}$ = (13h29m52.7s, +47d11m42.8s)
from \citet{Hagiwara:2001qy}.

\begin{figure*}
\epsscale{1.2}
\plotone{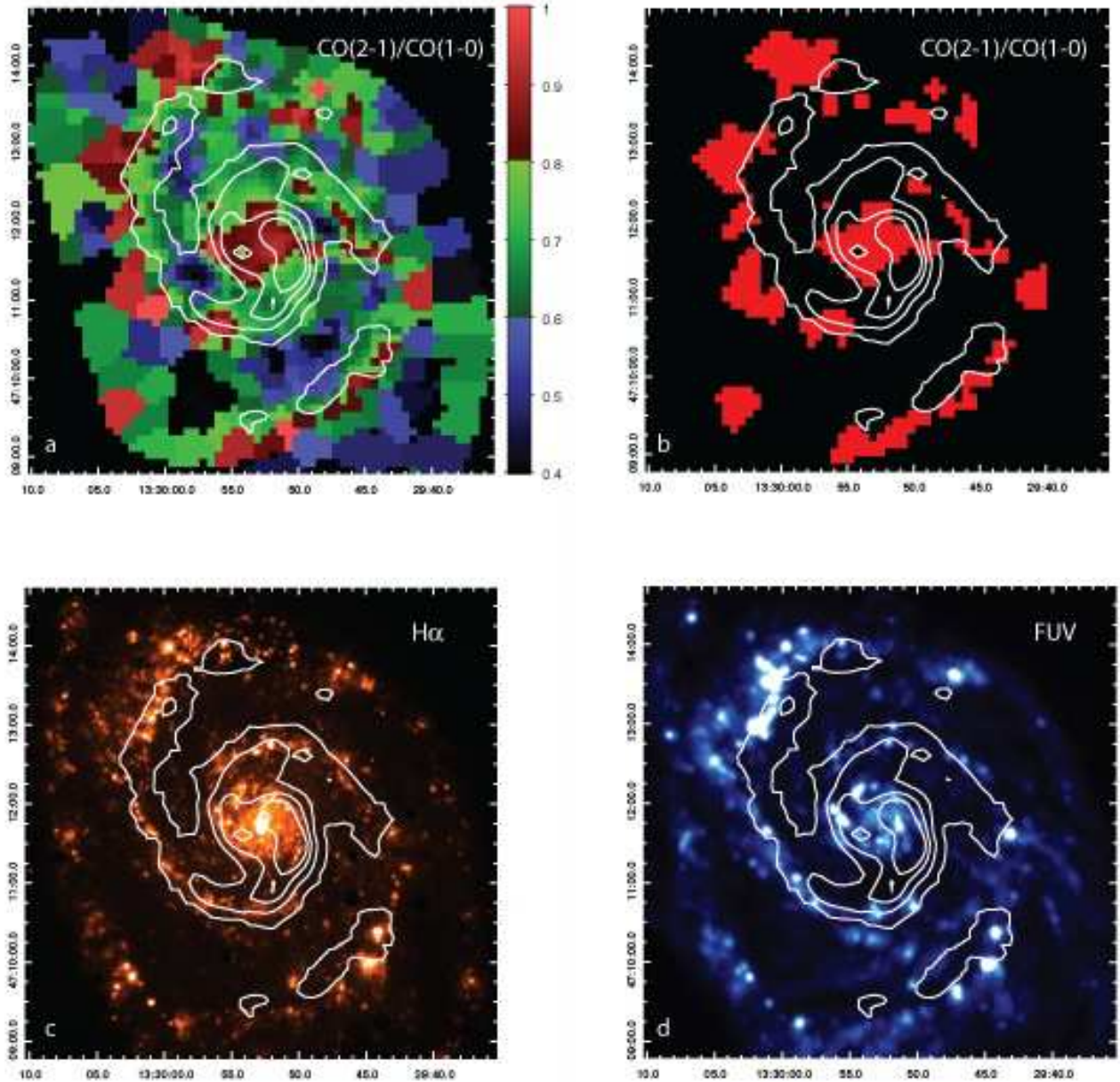}
\caption{
Locations of high and low $R_{2-1/1-0}$ with respect to the molecular spiral arms
and star forming regions.
(a) Map of $R_{2-1/1-0}$. The same as Figure \ref{fig:maps}c, but with a different color palette
to separate the gas with $R_{2-1/1-0}$ $<$0.6, 0.6-0.8, and $>$0.8.
This figure includes the outer areas where a large number of pixels ($>10$) need to
be averaged for reference.
(b) Map of high $R_{2-1/1-0}$ ($>0.8$).
(c) H$\alpha$ image.
(d) FUV image.
Contours are the same as those in Figure \ref{fig:maps} and are
at $I_{\rm CO(1-0)}=$12, 24, 36, and 48$\,\rm K\cdot \kmps$.
\label{fig:map2}}
\end{figure*}

\section{CO Line Ratios}\label{sec:ratio}

Line ratio images require high significance detections, especially of the denominator emission line;
therefore, in the outer disk of NGC5194
where the line brightness decreases,  smoothing and/or clipping of low significance data is required.
We adaptively smoothed the CO(1-0) data with the spatial binning method developed by \citet{Cappellari:2003yq}
and used data clipping in these areas when a large number of pixels ($>10$) need to be averaged;
identical smoothing and clipping was applied to the CO(2-1) map.
Figure \ref{fig:maps}a and b show the resultant maps for CO(1-0) and CO(2-1).

The noise levels change across the maps as a result of the variable velocity widths for integration and
adaptive smoothing, but the signal to noise ratios are high across the entire detected area,
including at the edges, i.e., $\gtrsim 10\sigma$ in both CO(1-0) and CO(2-1).

The 24$\micron$ image is processed in the same way and is shown in Figure \ref{fig:maps}d.
The 24$\micron$ emission appears at the leading (downstream) edges/sides of the CO(1-0) spiral arms (contours).
Figure \ref{fig:map2}c,d show the H$\alpha$ and FUV images at higher resolutions.
These emissions, tracing recent star formation, are in general at the leading edges of the molecular spiral arms.

\subsection{Spatial Distribution}\label{sec:spatial}

\begin{figure}
\epsscale{1.2}
\plotone{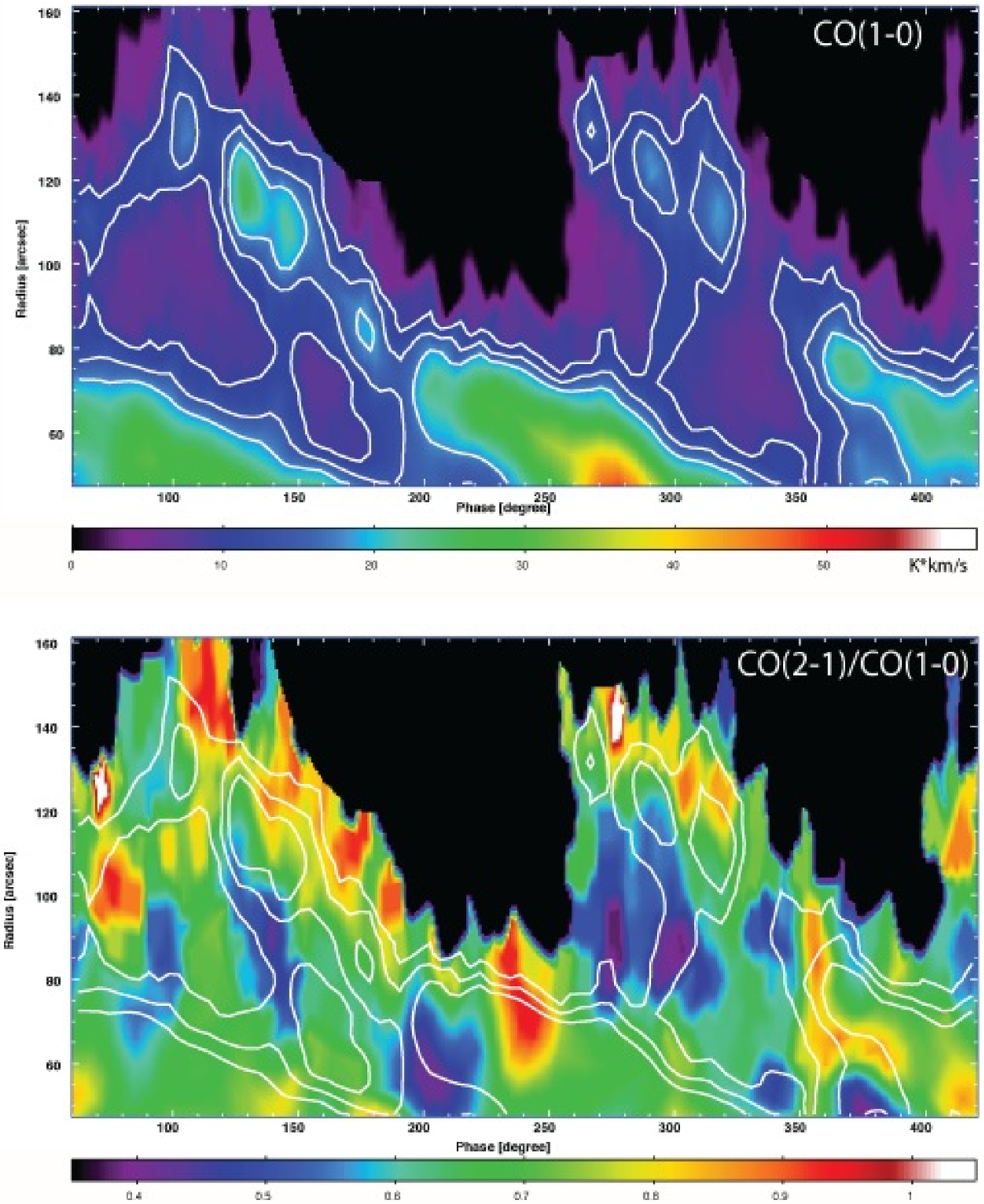}
\caption{Phase diagrams of the CO(1-0) integrated intensity (top panel) and $R_{2-1/1-0}$
with CO(1-0) contours (bottom). The azimuthal angle (phase) is defined counter-clockwise from the west.
The plot spans over the full $360\deg$, starting from $60\deg$ to cover the two spiral arms
without phase wraps. The counter-clockwise gas flow in Figure \ref{fig:maps} is toward right
in these diagrams. The gas with high $R_{2-1/1-0}$ appears predominantly at the downstream
edge/side of the molecular spiral arms.
\label{fig:phase}}
\end{figure}

Figure \ref{fig:maps}c is the $R_{2-1/1-0}$ map and shows a spiral pattern
of elevated $R_{2-1/1-0}$, i.e., $R_{2-1/1-0}$ varies {\it systematically} between the spiral arm and interarm regions.
In the interarm regions, this ratio is $<0.7$ and often as small as $\sim 0.4-0.6$,
while around the molecular spiral arms
it is elevated to  $>0.7$ and often becomes as high as $\sim 0.8-1.0$,
mainly at the leading (downstream) edges of the arms (see also Figure \ref{fig:map2}a, b).
The average $R_{2-1/1-0}$ over the entire disk is $\sim0.70$.
High ratios also appear occasionally at some upstream edges (e.g., in the north-west quadrant).
We note again that both CO(1-0) and CO(2-1) are detected significantly even at the edge of this map.
In the central $70\arcsec$ ($\sim 2.8$ kpc), $R_{2-1/1-0}$ is also high, $\sim 0.8-1.0$.
The high ratio appears to coincide with strong $24\mu m$ emission (Figure \ref{fig:maps}d).
Figure \ref{fig:phase} shows the phase diagrams of the CO(1-0) integrated intensity and
$R_{2-1/1-0}$, again evidencing the association of the high ratio with the molecular
spiral arms, especially at their leading edges/sides (i.e., right-hand side).

\begin{figure*}
\epsscale{1.1}
\plotone{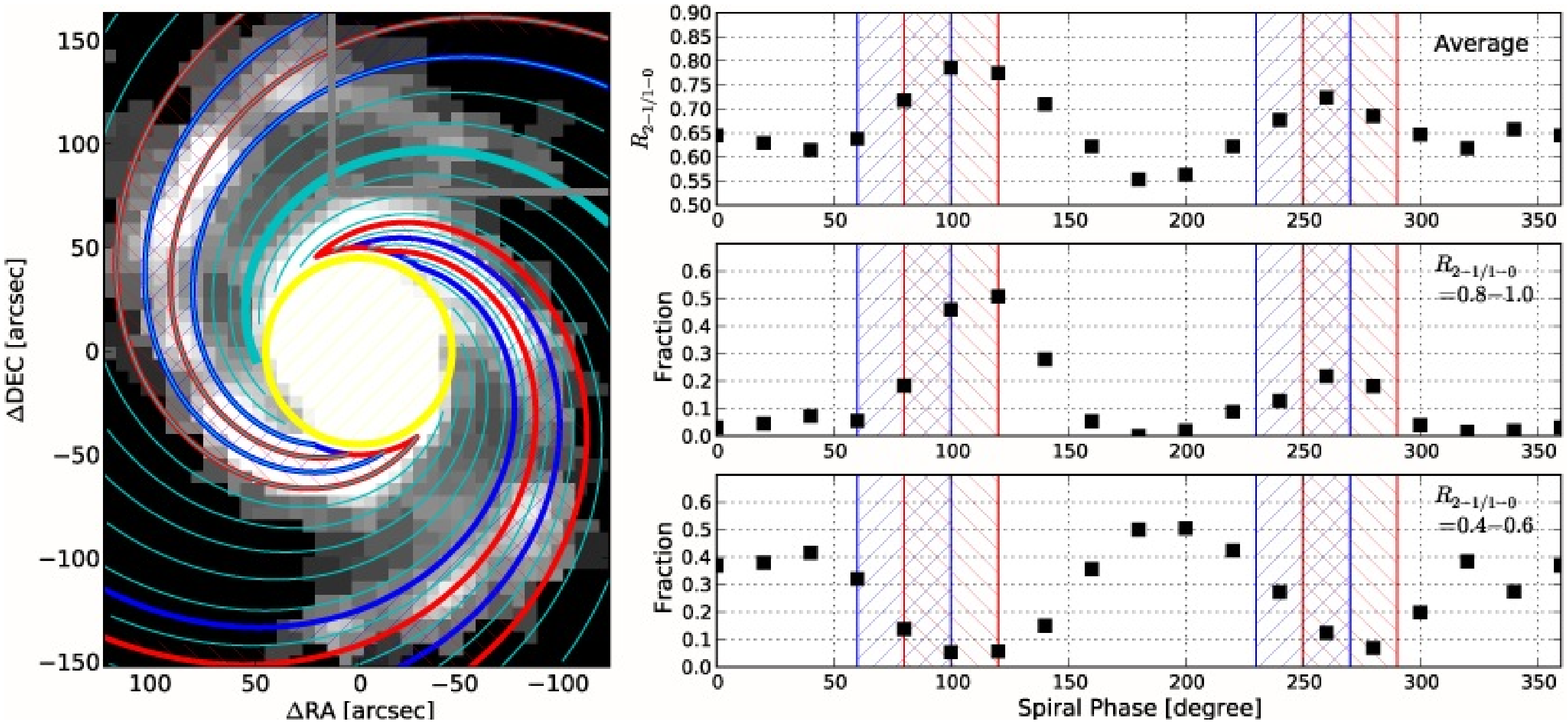}
\caption{Spatial variations of $R_{2-1/1-0}$.
{\it Left:} Definition of spiral arm phase ($\psi$). A logarithmic form of spiral pattern with
a pitch angle of $20\deg$ is adopted.
The spiral phase starts from the thick solid curve in cyan ($\psi = 0\deg$) and increases
counter-clockwise with thin cyan curves drawn at a $20\deg$ interval.
The background image is the same CO(1-0) image as Figure \ref{fig:maps}a.
Rough areas of two molecular spiral arms are hatched in blue ($\psi =$ 60-100
and $230$-$270\deg$). The leading edges of the molecular spiral arms
(and comparably, star-forming spiral arms in H$\alpha$ and FUV) are hatched in red
($\psi =$ $80$-$120$ and $250$-$290\deg$).
A molecular spiral arm deviates from this simple logarithmic form at the top-right corner
(indicated by gray solid lines), and we exclude this area from this analysis.
The central region is defined as the yellow circle with a radius of 45$\arcsec$.
{\it Right:} Average $R_{2-1/1-0}$ (top) and fractions of high ratio gas
(0.8-1.0; middle) and low ratio gas (0.4-0.6; bottom) as functions of the spiral phase.
Example histograms are presented in Appendix \ref{sec:hist}.
The typical standard deviation is $\sim 0.11$ in $R_{2-1/1-0}$ (top panel).
The Poisson errors in fractions are about $\sim 1\%$ (middle and bottom).
The values are calculated over $40\deg$ segments in $\psi$.
The molecular spiral arms and their leading (downstream) edges are hatched in blue and red, respectively,
as in the left image.
\label{fig:ratpi}}
\end{figure*}

The spatial variations of $R_{2-1/1-0}$ across spiral arms are also confirmed in Figure \ref{fig:ratpi}.
We adopt a simple logarithmic spiral pattern with a pitch angle of $20\deg$
and measure statistical quantities around the spiral curves over $40\deg$
segments ($20\deg$ increment).
The average $R_{2-1/1-0}$ increases to $>0.7$ in the molecular spiral arms (blue),
especially on/near their leading edges (red),
and decreases to $<0.7$ in the interarm regions.
The areal fraction of high ratio gas (0.8-1.0) is high in the spiral arms and their
leading edges, but low in the interarm regions.
The fraction of low ratio gas (0.4-0.6) is low in the arms,
but high in the interarm regions.
The central region also shows a high ratio. The average is 0.78 
within the central $90\arcsec$ ($\sim 3.6$ kpc) region and 0.82 within $70\arcsec$
($\sim 2.8$ kpc). [Example histograms of $R_{2-1/1-0}$ in the regions
are presented in Appendix \ref{sec:hist} and confirm the significant difference
among the regions.]

Figure \ref{fig:map2}a shows the distribution of gas with low ($R_{2-1/1-0}= 0.4$-$0.6$),
medium ($0.6$-$0.8$), and high ratios ($0.8$-$1.0$) with respect to the molecular spiral
arms (contours).
Figure \ref{fig:map2}b highlights the distribution of the high ratios ($0.8$-$1.0$).
For reference, this figure includes the outer areas where a large number of pixels ($>10$)
need to be averaged.
The variations of $R_{2-1/1-0}$ on large scales are evident:
high ratios around the molecular spiral arms (contours) as well as in the center,
and lower ratios in the interarm regions.
The high ratios appear mostly at the leading edge/side of the molecular spiral arms
where OB stars are predominantly located (Figure \ref{fig:map2}c, d).

\subsection{Correlations} \label{sec:corr}

\begin{figure*}
\epsscale{1.1}
\plotone{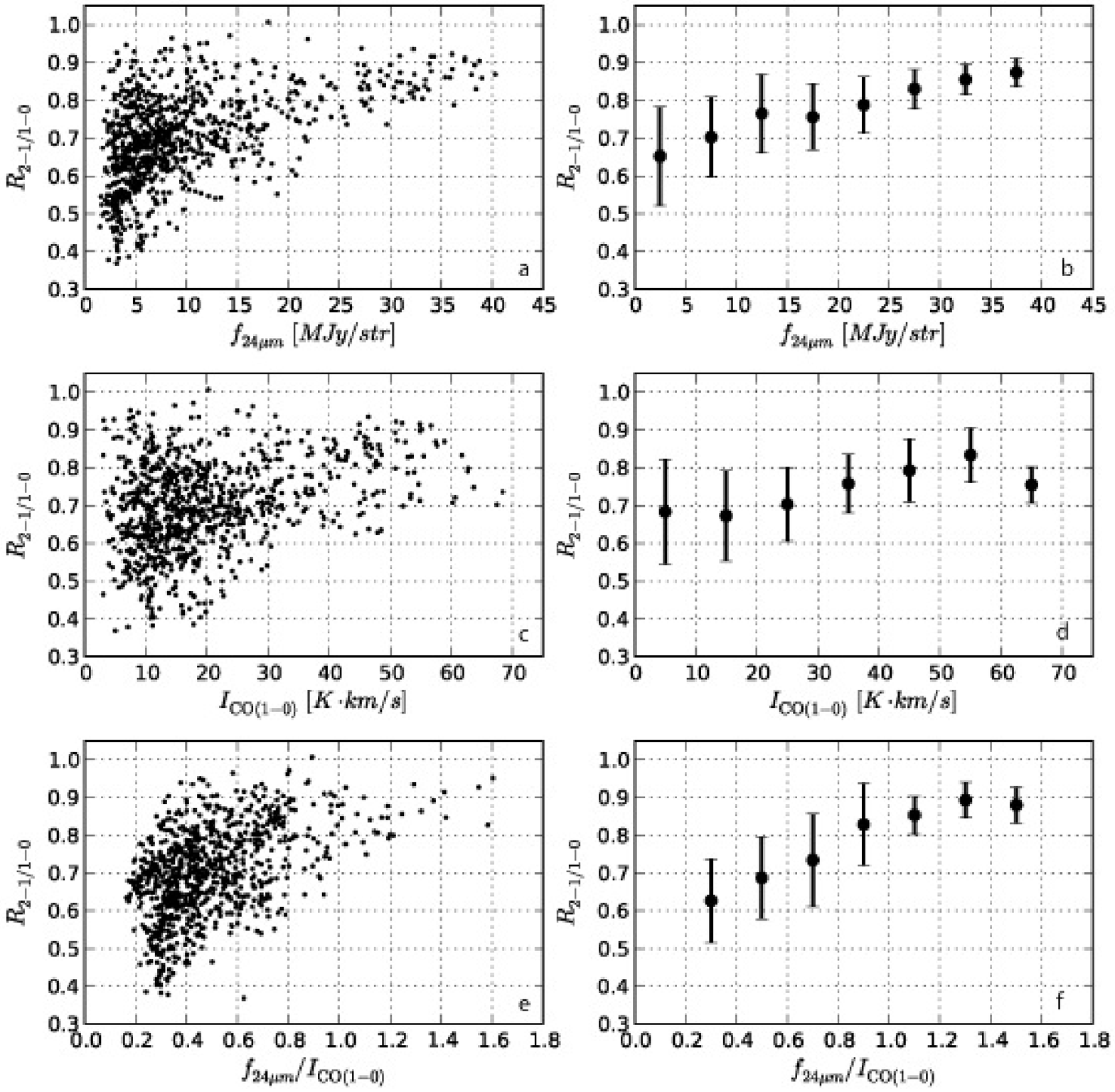}
\caption{
Correlations of the line ratio $R_{2-1/1-0}$ with
(a) the Spitzer 24$\mu m$ surface brightness $f_{24\mu m}$,
(c) CO(1-0) integrated intensity $I_{\rm CO(1-0)}$, and
(e) the ratio of $f_{24\mu m}$ and $I_{\rm CO(1-0)}$.
(b), (d), and (f) are the same as (a), (c), and (e), respectively, but with binned averages and standard deviations.
The $f_{24\mu m}$, $I_{\rm CO(1-0)}$, and $f_{24\mu m}/I_{\rm CO(1-0)}$ are approximately proportional to
the surface densities of star formation rate, molecular gas surface density, and star formation efficiency, respectively.
Note that there are some dependences between $x$ and $y$ axes in panel (c),(d),(e), and (f), since
both axes include $I_{\rm CO(1-0)}$. However, the dependences do not explain the absence of data
in the lower-right corners, since the detected parameter ranges can cover the corners.
\label{fig:ratio}}
\end{figure*}

Figure \ref{fig:ratio} shows pixel-by-pixel correlations of $R_{2-1/1-0}$ with three parameters
(for the pixels in Figure \ref{fig:maps}).
The $24\mu$m surface brightness $f_{24\mu m}$ is roughly a proportional of the star formation
rate surface density \citep{Calzetti:2007lr} and the CO(1-0) integrated intensity $I_{\rm CO(1-0)}$
is proportional to the molecular gas surface density.
Their ratio $f_{24\mu m}/I_{\rm CO(1-0)}$ is a measure of the star formation efficiency.
[Note that \citet{Calzetti:2007lr} discussed that the combination of $24\mu m$ and H$\alpha$ emission
provides a more accurate star formation rate; however, $f_{24\mu m}$ is used here as an approximate tracer,
since it shows inherent correlations. Correlations with some other star formation rate tracers
are given in Appendix \ref{sec:sfr}.]

Most remarkable are the lower envelopes, or cutoffs, of the distributions
in Figure \ref{fig:ratio} (left panels). No data points exist in the lower-right corners.
We note that there are some dependences between $x$ and $y$ axes in the middle and
bottom panels (though not in the top panels) as both axes include $I_{\rm CO(1-0)}$.
However, we conclude that the cutoffs in all panels are intrinsic to physical
conditions of the gas, since our detection limits cover the entire areas in these plots.
Hence, $R_{2-1/1-0}$ increases with the intensity of star formation and gas surface density.

The right panels show correlations using
averages and their scatter.
$R_{2-1/1-0}$ increases clearly with star formation activities (i.e.,
$f_{24\mu m}$ and $f_{24\mu m}/I_{\rm CO(1-0)}$) and weakly with
the gas surface density ($I_{\rm CO(1-0)}$).
The weaker correlation is due to the spatial offset between the molecular
spiral arm and high $R_{2-1/1-0}$ (\S \ref{sec:spatial}).
Multiple physical processes may contribute to the intrinsic scatters
in the left panels, and therefore, linear model fits may be inappropriate.
For instance, $R_{2-1/1-0}$ can be enhanced due to
increases in density prior to star formation and
increases in gas temperature due to heating after star formation (\S \ref{sec:discussion}).

\begin{figure*}
\epsscale{1.1}
\plotone{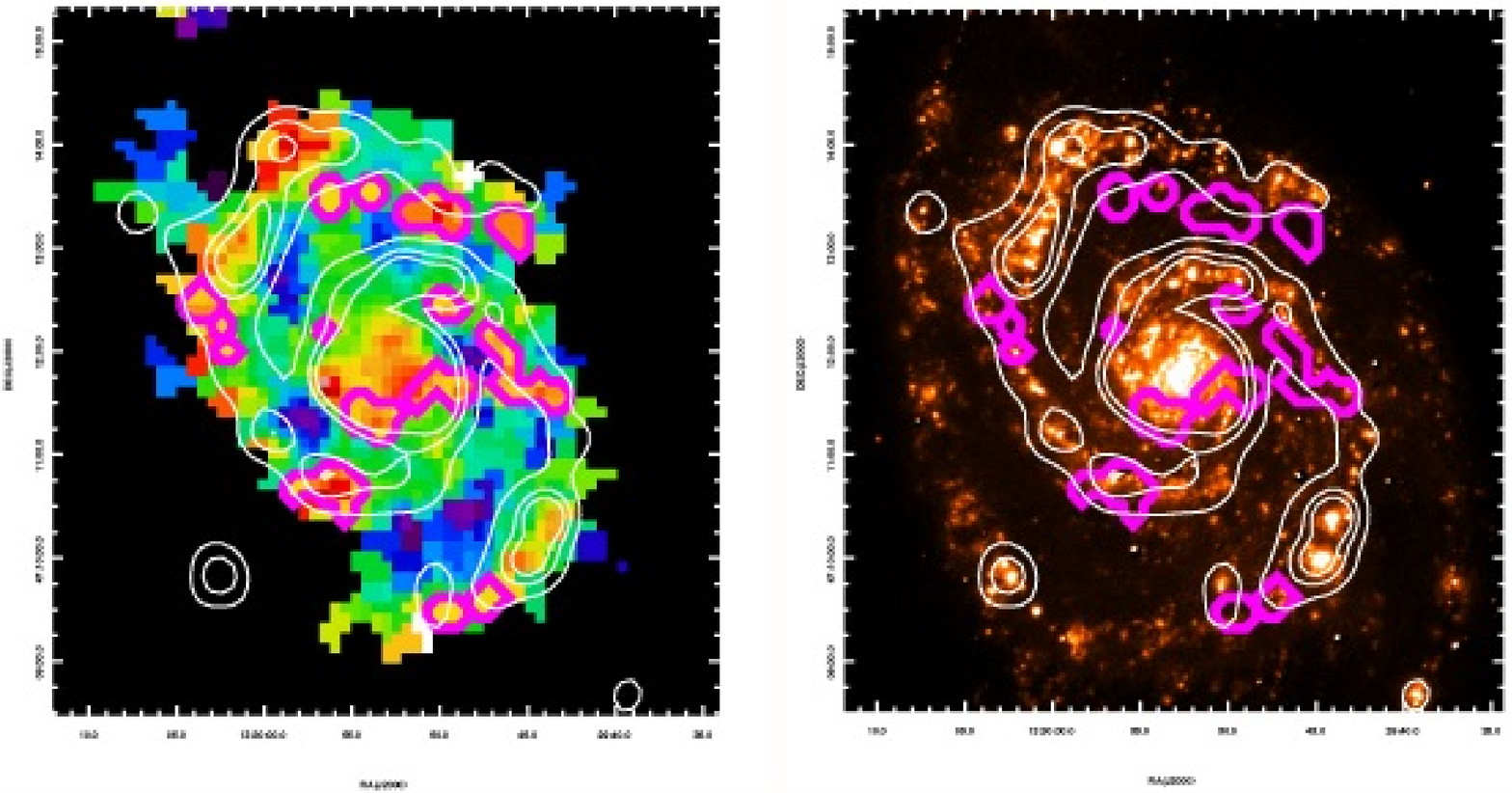}
\caption{Comparisons of the CO $J=2-1/1-0$ line ratio $R_{2-1/1-0}$ (left), H$\alpha$ image (right), and
contours (white) from the Spitzer 24$\mu$m image (without smoothing) at 4, 8, and 12 MJy/str.
The color scale of the left panel is the same as that in Figure \ref{fig:maps}.
The high $R_{2-1/1-0}$ regions coincide mostly with star forming regions (with high $f_{24\mu m}$ and HII regions). 
The contours in magenta are the regions of high $R_{2-1/1-0}$, $>0.8$, and low star formation efficiency,
$f_{24\mu m}/I_{\rm CO(1-0)}<0.6$ (see Figure \ref{fig:ratio}). These regions appear both on the spiral arms
and their trailing/upstream edges (e.g., in the north-west quadrant), possibly indicating
the growth of pre-star forming dense cores.
\label{fig:comp}}
\end{figure*}

In addition to the clear correlations with star formation activities,
some high values of $R_{2-1/1-0}$ are seen with
low $f_{24\mu m}$, $I_{\rm CO(1-0)}$, and $f_{24\mu m}/I_{\rm CO(1-0)}$.
Figure \ref{fig:comp} shows the locations of this component, i.e.,
with a high ratio, $R_{2-1/1-0} > 0.8$ and low star formation efficiency,
$f_{24\mu m}/I_{\rm CO(1-0)}<0.6$, with respect to $f_{\rm 24\mu m}$ (white contours)
and H$\alpha$ emission (right panel). 
Some regions with the elevated ratios appear at the upstream edges of the $24\mu$m spiral arm,
possibly indicating dense pre-OB star formation gas (see Section \ref{sec:discussion}).

\section{Discussion}\label{sec:discussion}

GMCs are abundant even in the interarm regions in M51 \citep{Koda:2009wd}.
Although our $19.7\arcsec$ ($\sim$780 pc) beam  typically contains many GMCs,
we adopt a one-zone approximation to obtain insights into
the mean physical conditions of the molecular gas and its spatial variations.
That is, we assume that the GMCs within a resolution element share the same average properties
and that CO(1-0) and CO(2-1) are emitted from the same physical regions.

\subsection{Large Velocity Gradient (LVG) Analysis} \label{sec:lvg}

The Large Velocity Gradient (LVG) approximation is often used for the line radiation transfer 
of interstellar molecular lines  \citep{Scoville:1974yu, Goldreich:1974jh}.
This approach enables treatment of the coupled radiative transfer and molecular excitation as a local problem 
since the line emission in different volumes of gas is doppler shifted and decoupled; the approach is 
justified by the fact that thermal line
broadening ($\sim 0.2$-$0.3 \kmps$) is much smaller than velocity dispersions within
GMCs ($\gtrsim$ a few $\kmps$).
The molecular excitation conditions then depend on the local kinetic
temperature $T_{\rm kin}$, volume density $n_{H_2}$ (since low-$J$ CO
emission is excited collisionally) and on the local line opacity (i.e.,
CO column density $N_{\rm CO}$ per unit velocity change).
Thus, ($T_{\rm kin}$, $n_{H_2}$, $N_{\rm CO}/dv$) determine the line ratio $R_{2-1/1-0}$, or conversely,
an observed $R_{2-1/1-0}$ constrains these parameters, albeit with some degeneracy.
We have developed our own LVG calculation code using the newest CO-H$_2$ collisional cross
sections from \cite{Yang:2010fk}.

The CO column density per unit velocity gradient is typically in the range 
$\log(N_{\rm CO}/dv\, [\rm cm^{-2} (\kmps)^{-1}])=16.6-17.3$ based on the observed line widths and 
assuming a standard ISM CO/H$_2$ abundance of $8\times10^{-5}$ \citep{Schinnerer:2010kr}.
For example, Galactic GMCs have approximately constant mean surface densities
averaged over their entire areas \citep[$\sim 170 \, \Msun \rm pc^{-2}$; ][]{Solomon:1987pr, Rodriguez-Fernandez:2006kl} and a range of
velocity widths $\sim 4$-$18 \kmps$, resulting in the above range of $N_{\rm CO}/dv$.
For the most  extreme surface densities in M51's spiral arms, the mass surface density is $\sim 1000\,\Msun \rm pc^{-2}$,
while  the velocity width is 50-100 $\kmps$ \citep{Koda:2009wd}, yielding a similar $N_{\rm CO}/dv$.

\begin{figure*}
\epsscale{0.8}
\plotone{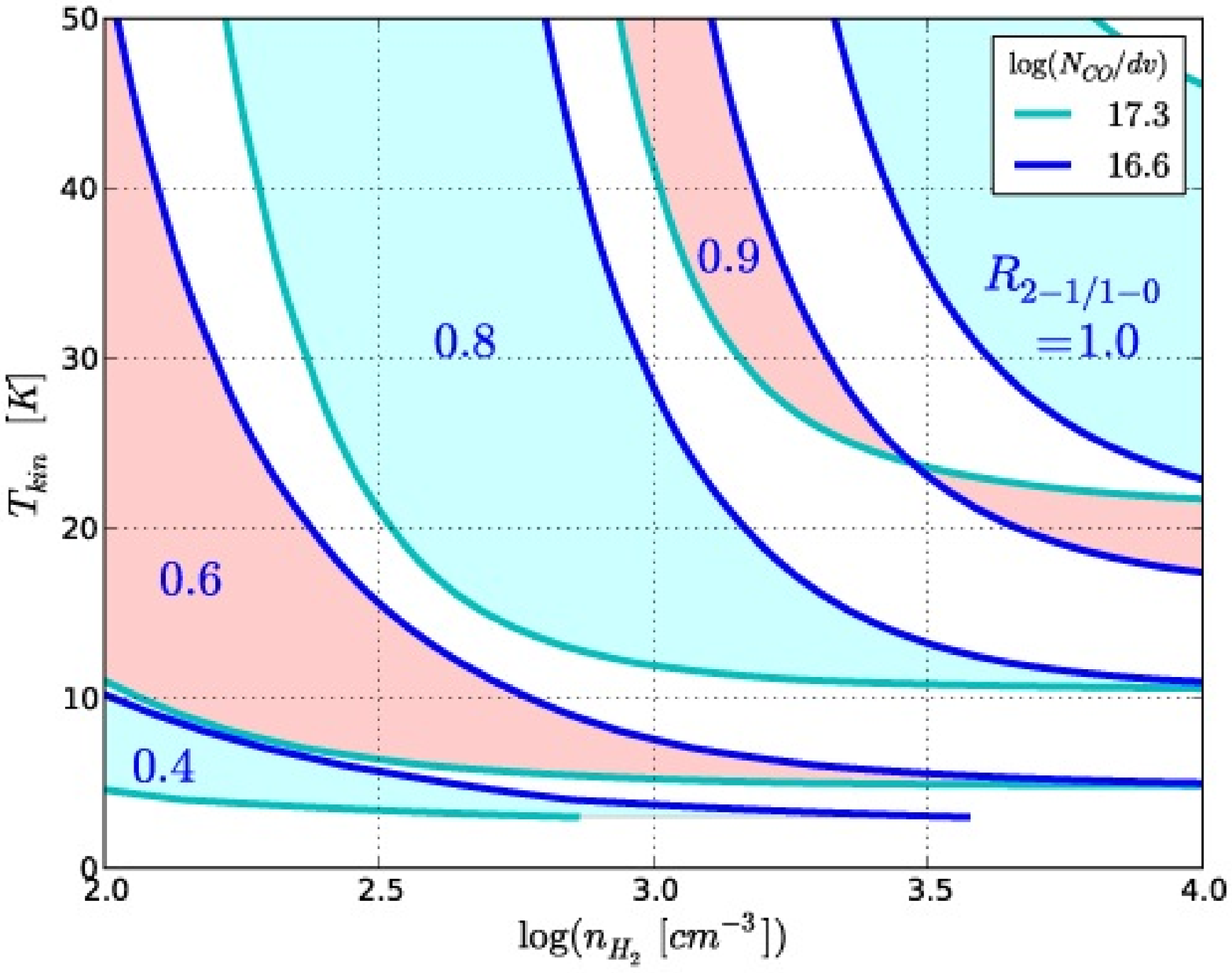}
\caption{
Large Velocity Gradient (LVG) calculations.
The line ratios $R_{2-1/1-0}$ are plotted as functions of
the gas kinetic temperature $T_{\rm kin}$ and $\rm H_2$ density $n_{\rm H_2}$.
Most GMCs are in the range of the CO column densities
$\log(N_{\rm CO}/dv)=16.6$ (blue) and $17.3$ (cyan) assuming the
CO fractional abundance to H$_2$ of $8\times 10^{-5}$.
The regions between the curves for the two column densities (blue and cyan) are shaded for clarity.
\label{fig:lvg}}
\end{figure*}

Figure \ref{fig:lvg} shows the parameter ranges for $T_{\rm kin}$ and $n_{H_2}$ which yield 
values of $R_{2-1/1-0} = 0.4, 0.6, 0.8, 0.9$ and $1.0$. In each case, two curves are shown corresponding 
to $\log(N_{\rm CO}/dv)=17.3$ and $16.6$. (Note that 
the relative position of the two curves is reversed for $R_{2-1/1-0} \geq 0.9$, compared to that for lower values of $R_{2-1/1-0}$.)
Values of $R_{2-1/1-0} < 0.4$ and $>1$ can be obtained but are not shown here since they are not seen in our observations.
Galactic GMCs typically have $T_{\rm kin}\sim 10$ K and $n_{H_2}\sim 300 \rm\, cm^{-3}$
($\log n_{H_2}\sim 2.5$) with a scatter of about a factor of 2 \citep{Scoville:1987vo,Solomon:1987pr}, yielding
a range in $R_{2-1/1-0}\sim 0.4$-$0.8$ and a mean observed ratio of $0.6$-$0.7$ from Figure \ref{fig:lvg}.
This range is consistent with the observed $R_{2-1/1-0}$ in interarm GMCs in the Galaxy and
the bulk parts of the Orion GMCs, i.e., the larger region not closely associated with active high mass star formation \citep{Sakamoto:1997ys}.

Figure \ref{fig:lvg} shows that $R_{2-1/1-0}$ is a sensitive probe of physical conditions
in the {\it bulk} molecular gas.
A high value of $R_{2-1/1-0} \sim 0.8$ requires increases in either $T_{\rm kin}$
or $n_{H_2}$ (or both) by a factor of $\sim$2-3, relative to the standard GMC
values of 10 K and 300 cm$^{-3}$.
To reproduce the even higher ratios $\sim0.9$-$1.0$ within an expected range of
$T_{\rm kin}$ or $n_{H_2}$, $T_{\rm kin}$ and $n_{H_2}$  need to increase
more than factors of $\sim$3 and $\sim$10, respectively.

$R_{2-1/1-0}$ shows spatial variations in M51. 
The ratio often reaches as low as $\sim 0.4$-$0.6$ in the interarm regions,
while it becomes as high as $\sim 0.8$-$1.0$ in the spiral arms
(Figure \ref{fig:maps}, \ref{fig:map2}, and \ref{fig:ratpi}).
The lower $R_{2-1/1-0}$ ($\sim$0.4-0.6) generally seen in the interarm regions implies that 
the gas there has significantly lower $T_{\rm kin}$ and $n_{\rm H_2}$ than that in the arms.
Based on the similarity in  $R_{2-1/1-0}$, we conclude that the GMCs in M51's interarm regions 
have typical physical conditions similar to their Galactic counterparts.
In the Galaxy, the mass dividing line for 50\% of the total H$_2$ content is $\sim 4\times 10^5\, \Msun$, i.e. GMCs more massive than this 
account for half the total molecular gas mass \citep{Scoville:1987vo}.
In M51, we can derive a similar GMC mass 50\% percentile of $\sim5 \times10^5\, \Msun$ using
higher resolution interferometric imaging \citep{Koda:2009wd,Koda:2011nx} which is capable of resolving 
the emission of more massive clouds and comparing their total emission with that in the single dish maps which also include 
contributions from lower mass clouds.

The higher $R_{2-1/1-0} \sim$0.8-1.0, seen typically on the leading (downstream) edges
of the spiral arms,
requires higher values, both in $T_{\rm kin}$ and $n_{\rm H_2}$ by a factor of 2-3,
compared to those of typical GMCs and interarm GMCs.
This downstream gas is associated with ongoing star formation (Section \ref{sec:ratio}).
Therefore, increases in both $T_{\rm kin}$ and $n_{\rm H_2}$, rather than 
in either $T_{\rm kin}$ or $n_{\rm H_2}$ alone, may be more likely.
On the other hand, the gas with high $R_{2-1/1-0} \sim$0.8-1.0 on the upstream edge
typically shows low star formation activity (Figure \ref{fig:comp}), possibly explained by increases
in $n_{\rm H_2}$ alone, without significant increase in $T_{\rm kin}$.

Of course, this one-zone analysis is simplistic for regions extended over $\sim 780$ pc,
but it still provides an indication that the bulk properties of the gas
evolve from the interarm regions through the spiral arms.
{\it The interarm gas is mostly in GMCs without active star formation,
and upon entry into the spiral arms, the molecular gas becomes warmer and/or denser
by factors of at least $\sim 2-3$, possibly due to dynamical compression
(e.g., at the upstream edges), but primarily due to young stellar heating (at the downstream edges).}

\subsection{Comparison with Galactic Studies}\label{sec:mw}

A similar evolution of molecular gas, i.e., warmer and/or denser in spiral arms, has been
observed in the Galaxy, although uncertainties remain due to our in-plane viewpoint for the Milky Way.
\citet{Sanders:1985ud} found GMCs with high brightness temperature
mostly associated with spiral arms as delineated by radio HII regions.
\citet{Sawada:2012fk, Sawada:2012lr} recently demonstrated that the molecular gas with high brightness temperature is located continuously along the spiral arms in the longitude-velocity diagram.
\citet{Sakamoto:1995lr, Sakamoto:1997ys} found, from their Galactic Plane CO(2-1) survey,
that molecular gas with the highest $R_{2-1/1-0}$ is predominantly along the Sagittarius
and Scutum spiral arms. Both the Galaxy and M51 exhibit similar variation in $R_{2-1/1-0}$
across their disks.

The great advantage of the Galactic studies is their high linear resolution.
\citet{Sakamoto:1997ys} and \citet{Hasegawa:1997lr} classified molecular gas
into three categories based on their $R_{2-1/1-0}$ measurements:
low ratio gas ($R_{2-1/1-0}<0.7$), high ratio gas ($0.7$-$1.0$), and very high ratio gas ($>1.0$).
The low ratio gas is often observed in large, non-star forming
parts of GMCs \citep{Sakamoto:1994lr} and in interarm GMCs \citep{Sakamoto:1997ys}.
The very high ratio gas appears exclusively immediately around HII regions
\citep[e.g., the M42 nebula; ][]{Sakamoto:1994lr}, and hence is likely strongly heated
by radiation at the boundary of the HII regions.
The high ratio gas can also be found around the HII regions, just outside the very high ratio gas.
In Galactic GMCs, high ratios are also associated with dense core regions prior to efficient star formation, but where 
the CO levels have become fully thermalized and the temperatures are modestly elevated \citep{Hasegawa:1997lr}.

The central $\sim 2.8$ kpc region of M51 shows high $R_{2-1/1-0} (\sim 0.8$-$1.0$).
This is similar to the Galactic center, where $R_{2-1/1-0} \sim 1.0$, and sometimes $>1$,
over its central 900 pc region. These ratios are quite high in comparison
with typical GMC values, though they are around the average for galactic centers
\citep[some starburst galaxies show $R_{2-1/1-0} >1.0$ in their centers; ][]{Braine:1992lr, Sawada:2001lr}.
\citet{Sawada:2001lr} explain that the high and very high ratios in the Galactic center
are caused by intrinsically high $T_{\rm kin}$ and $n_{\rm H_2}$, combined with low opacity
in low $J$ transitions due to the large velocity dispersion and a very high excitation ($J\geq3$).
The high ratios ($0.8$-$1.0$) in the center of M51 probably indicate similar physical
conditions to those of the Galactic Central Molecular Zone.

\subsection{Causes of High Ratios}

The spiral arm and galactic nucleus GMCs with higher $R_{2-1/1-0}$ are likely heated primarily by associated OB star formation (Figure \ref{fig:maps}, \ref{fig:map2}, and \ref{fig:ratio}),
and their densities raised by dynamical compression within the spiral arms (i.e., close-encounters and/or
collisions with other GMCs) and shearing \citep{Koda:2009wd},
all of which can lead to the high ratio gas ($>0.7$; and up to 0.8-1.0).

A substantial fraction of ionizing photons leaking out of HII regions \citep{Ferguson:1996qy, Wang:1999uq, Blanc:2009ys}
might be a cause of the elevated heating outside the HII regions.
However, to alter the  temperatures over the very large regions ($\gtrsim 780$ pc) at the leading sides of the spiral arms (as seen here),
it seems more likely that the radiative heating is provided by the longer wavelength continuum of the young star clusters
(i.e., longward of the Lyman limit) combined with hydrogen recombination line radiation
from the HII regions (since the Lyman continuum is almost surely absorbed over these long paths). 
An abundant population of active HII regions exists along the spiral arms, some of them
10-100 times brighter than the Orion nebula \citep{Scoville:2001qm} and of course there is considerable luminosity 
longward of the Lyman limit.
High ratios over a large area are also found around 30 Doradus complex
in the Large Magellanic Cloud, one of the most active star forming regions in the Local Group \citep{Sorai:2001lr}.

We note that cloud-cloud tidal interactions within the spiral arms
can also inject a large amount of internal energy into GMCs in very close encounters.
The typical gas thermal energy in a GMC is $\sim 2\times 10^{47}$ ergs; this is two orders of magnitude smaller than
the energy increase due to a close cloud-cloud encounter, $\sim 3\times 10^{49}$ erg [estimated using the impulse approximation
\citep{Spitzer:1958lr}, an encounter speed of $10\kmps$, and typical GMC parameters \citep{Scoville:1987vo}].
Dynamical environments in the spiral arms can possibly explain high ratios
at the spiral arm entry (at their upstream edges).

\section{Implications to Other Studies}
Our results indicate that CO(2-1) emission is sensitive to the change of physical conditions
of bulk molecular gas in GMCs, in contrast to CO(1-0), which traces the bulk molecular mass.
The spatial variation of $R_{2-1/1-0}$ indicates the evolution in physical conditions of the gas
as it passes from interarm to arm regions during the course of azimuthal orbital motion.
The use of CO(2-1) is reasonable as a tracer of molecular gas mass when galactic structures are unresolved,
e.g., in studies of high redshift galaxies \citep[e.g., ][]{Frayer:2008qe},
since the $R_{2-1/1-0}$ varies only by a factor of $\sim2$.
However, spatially-resolved studies of CO (2-1) in nearby galaxies may be quite misleading with respect 
to inferred variations in the gas mass surface density due to changes in 
$R_{2-1/1-0}$ with both radius (e.g. in the nuclear regions) and with azimuth due to the spiral arms.
The systematic trends illustrated in Figure \ref{fig:ratio} will affect correlation studies
of the gas mass and other parameters such as the Schmidt relation \citep{Bigiel:2008kx,
Leroy:2008mq, Schruba:2011lr}. In fact, the Schmidt relation using the CANON CO(1-0) data
of $\sim$10 galaxies
shows a non-linear, power-law relationship \citep[][Momose et al. in prep.]{Liu:2011pd},
as opposed to the linear relation derived from the above quoted CO(2-1) studies.

\section{Summary}
New multi-beam receivers and OTF mapping technique in millimeter astronomy enable us to map
the large molecular disk of M51 in CO $J=1-0$ and $2-1$ transitions with unprecedented accuracy.
We find systematic spatial variations in the line ratio $R_{2-1/1-0}$ -- from low ($<0.7$; often
$\sim 0.4-0.6$) in the interarm regions to high ($>0.7$; $\sim 0.8-1.0$) in the spiral arms,
especially on their leading (downstream) edge.
The central $\sim 2.8$ kpc region also exhibits high $R_{2-1/1-0}$ ($\sim 0.8-1.0$). 

The interarm gas is mostly in GMCs without
active star formation, and upon entry into the spiral arms, the molecular gas becomes warmer and/or denser
by factors of $\sim 2-3$. The gas in the spiral arms and galactic nucleus are likely
heated primarily by associated OB star formation (especially their continuum radiation longward
of the Lyman limit) and their densities raised by dynamical compression within the spiral arms.
Some high ratio gas exists on the upstream edges of the spiral arms, perhaps indicating
gas becoming dense prior to star formation.
Similar variations in the CO line ratios have been seen in the Galaxy -- GMCs with inactive star formation
have low $R_{2-1/1-0}$  ($\sim 0.4-0.6$) and ones with active star formation show
high $R_{2-1/1-0}$ ($\sim 0.8-1.0$).

\acknowledgments

The Nobeyama 45-m telescope is operated by the Nobeyama Radio Observatory, a branch of the National Astronomical Observatory of Japan.
We thank Seiichi Sakamoto and Jim Barrett for valuable comments. We also thank the anonymous referee for valuable comments.
JK acknowledges support from the NSF through grant AST-1211680 and NASA through grant NNX09AF40G, an Hubble Space Telescope grant, and a Herschel Space Observatory grant.

\clearpage
\appendix

\section{Histograms of $R_{2-1/1-0}$}\label{sec:hist}
\begin{figure*}
\epsscale{1.1}
\plotone{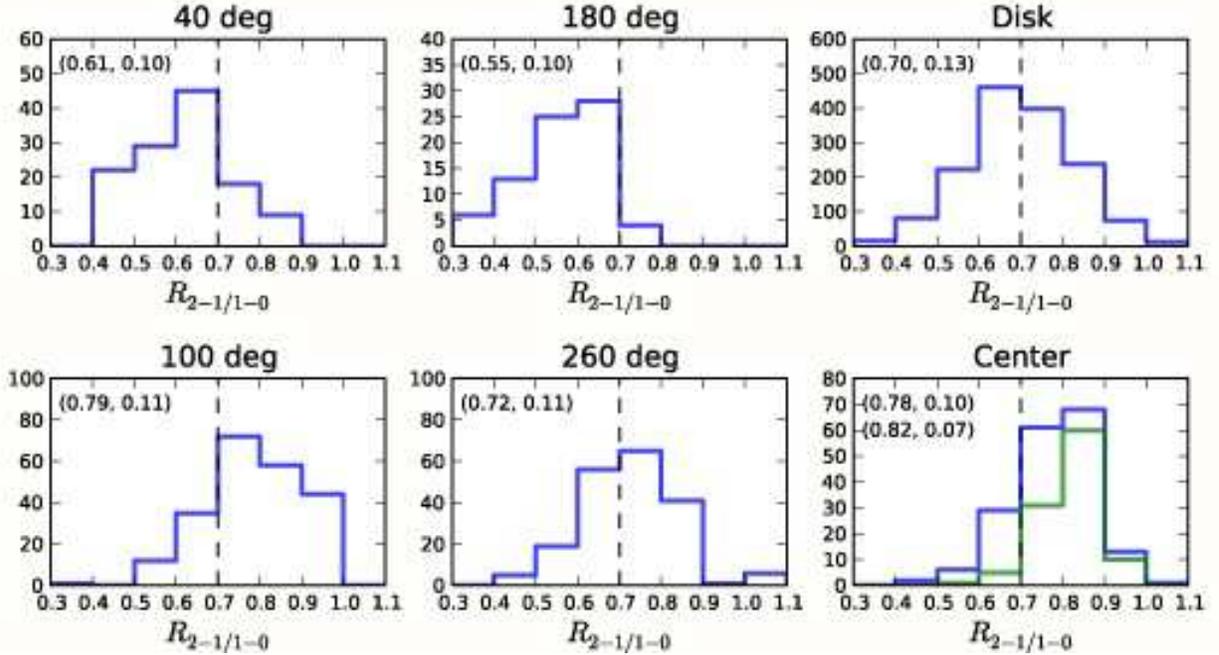}
\caption{Sample histograms of $R_{2-1/1-0}$ at spiral arm regions (i.e., the spiral phases of $100\deg$ and $260\deg$ in Figure \ref{fig:ratpi}), interarm regions ($40\deg$ and $180\deg$), the entire disk, and the central region. The average and standard deviation are in the parenthesis at the top-left corners. Two histograms are drawn for central regions with the diameters of $90\arcsec$ ($\sim 3.6$ kpc; blue) and $70\arcsec$ ($\sim 2.8$ kpc; green).
\label{fig:hist}}
\end{figure*}

Figure \ref{fig:hist} presents sample histograms of $R_{2-1/1-0}$ in spiral arms, interarm regions, entire disk, and central region defined in Figure \ref{fig:ratpi}. Differences in the average $R_{2-1/1-0}$ between the spiral arms and interarm regions are significant compared to the standard deviations.

The average ratio of the spiral arm of $260\deg$ is smaller than that of $100\deg$. The fraction of high ratio gas is also lower for the $260\deg$ arm (Figure \ref{fig:ratpi}), though the increase from interarm regions is still evident. These differences are partly due to the high ratio gas appearing upstream this spiral arm at smaller radii and downstream at larger radii (Figure \ref{fig:maps} and \ref{fig:map2}). A full discussion of the nature of this spatial variation (upstream vs downstream) is beyond the scope of this paper, but possible causes include the scattered distribution of star forming regions along this arm \citep{Egusa:2009bv}, cloud-cloud interactions, or dynamical gas compression due to the wiggle instability \citep{Wada:2004fu}.

\section{Correlations between $R_{2-1/1-0}$ and Star Formation Rate Tracers}\label{sec:sfr}

\begin{figure*}
\epsscale{1.2}
\plotone{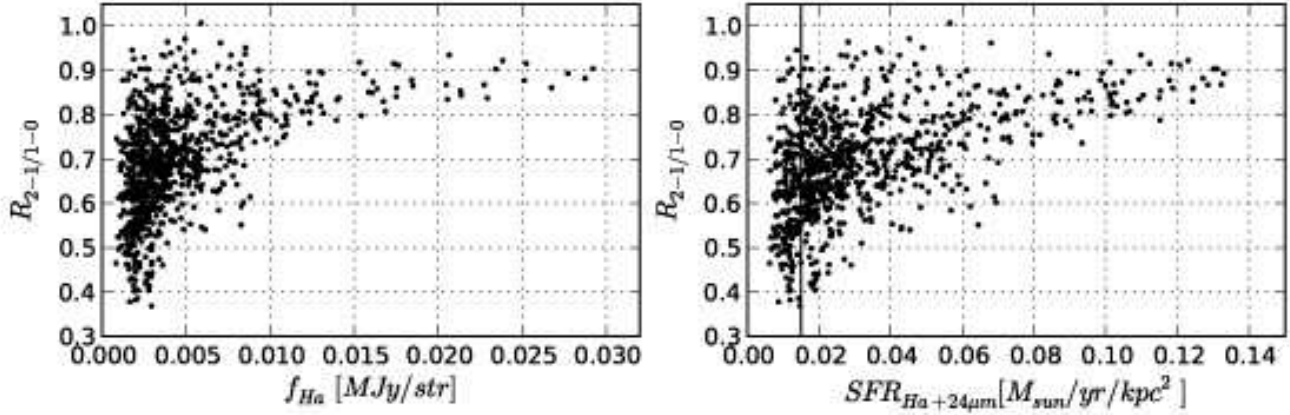}
\caption{$R_{2-1/1-0}$ as functions of (a) H$\alpha$ emission and (b) star formation rate (SFR) calculated with $24\mu m$
and H$\alpha$ emission \citep{Calzetti:2007lr, Kennicutt:2007vy}.
The solid vertical line in (b) indicates a boundary below which the background emission unrelated to recent star formation may contaminate the measurement \citep{Calzetti:2007lr, Liu:2011pd}.
\label{fig:sfr}}
\end{figure*}

We adopted the $24\micron$ image as an approximate tracer of star formation rate and showed the
correlation between $R_{2-1/1-0}$ and star formation activities in \S \ref{sec:corr}.
Similar correlations exist with other tracers of star formation rate.
Figure \ref{fig:sfr} shows correlations with H$\alpha$ flux and with star formation rate
calculated by combining 24$\micron$ and H$\alpha$  \citep{Calzetti:2007lr, Kennicutt:2007vy}.

\end{document}